\documentclass{optica-article}

\journal{opticajournal}
\articletype{Research Letter}
\usepackage{lineno}
\usepackage{siunitx}
\usepackage{amsmath}
\makeatletter
\fancypagestyle{plain}{%
  \fancyhf{}%
  \fancyfoot[C]{\thepage}%
}
\renewenvironment{abstract}
{\vskip1pc\noindent\textbf{Abstract:\space}}
{\par\vskip12pt}
\makeatother

\usepackage{xcolor}

\usepackage{cleveref}
\crefname{equation}{equation}{equations}
\Crefname{equation}{Equation}{Equations}
\crefname{figure}{figure}{figures}
\Crefname{figure}{Figure}{Figures}

\DeclareUnicodeCharacter{2009}{\,}

\newcommand{\FP}{Fabry-P\'{e}rot}

\definecolor{magenta}{rgb}{0.8,0,0.8}

\definecolor{darkgreen}{rgb}{0,0.7,0}

\newcommand{\be}{\begin{equation}}
\newcommand{\ee}{\end{equation}}
\newcommand{\bea}{\begin{eqnarray}}
\newcommand{\eea}{\end{eqnarray}}

\newcommand{\Ei}{{\rm Ei}}

\newcommand{\Imr}{{\rm Im}}

\newcommand{\vb}[1]{\mathbf{#1}} % vector bold style

\begin{document}

\title{A thermoelastic limit on the focal intensity in \FP{} cavities}

\author{Jeremy J. Axelrod\authormark{1,3}, Lothar Maisenbacher\authormark{1}, Ashwin Singh\authormark{1}, Isaac M. Pope\authormark{1}, Petar N. Petrov\authormark{1}, Jessie T. Zhang\authormark{1},  Holger M\"{u}ller\authormark{1,2,*}}

\address{\authormark{1}Department of Physics, University of California, Berkeley, Berkeley, USA\\
\authormark{2}Lawrence Berkeley National Laboratory, Berkeley, USA\\
\authormark{3}Department of Molecular and Cellular Physiology, Stanford University, Stanford, USA}

\email{\authormark{*}hm@berkeley.edu} %% email address is required; see note below about the corresponding author designation



\begin{abstract}
Light in the mode of a \FP{} cavity heats the mirror surfaces via optical absorption, causing thermoelastic deformation of the mirror substrates, which in turn dictates the shape of the mode. We develop an analytical model which predicts that this effect limits the maximum focal intensity of the mode. Using two near-concentric \FP{} cavities---one with $4.5$-fold higher mirror absorption than the other---we measure the thermoelastic properties of the cavity mirrors and demonstrate that it is possible to achieve at least $70\%$ of this predicted limit (in the high-absorption cavity), and that the predicted limit is $\qty{2.9}{\tera\watt\per\centi\meter^2}$ (in the low-absorption cavity).
\end{abstract}

%%%%%%%%%%%%%%%%%%%%%%%%%%  body  %%%%%%%%%%%%%%%%%%%%%%%%%%
\section{Introduction}
\FP{} optical cavities are widely used in optical science and technologies including lasers\cite{siegmanLasers1986}, telecommunication\cite{venghausWavelengthFilters2017,mallinsonWavelengthselectiveFiltersSinglemode1987}, atomic physics\cite{alvarezOpticalCavitiesOptical2019}, spectroscopy\cite{berdenCavityRingdownSpectroscopy2000,gianfraniAdvancesCavityenhancedMethods2024}, quantum computation/information\cite{reisererCavitybasedQuantumNetworks2015,  giovannettiScalableQuantumComputation2000}, and gravitational wave detectors\cite{collaborationAdvancedLIGO2015,tseQuantumEnhancedAdvancedLIGO2019}. The shape of the cavity mirrors and the length of the cavity determine the spatial and spectral properties of the cavity's electromagnetic modes. It is well-known that heating of the mirrors by the cavity mode results in thermoelastic deformation of the mirror surface, thereby changing the properties of the cavity. For example, this may create unwanted degeneracy between modes\cite{bullingtonModalFrequencyDegeneracy2008}, change the cavity's input coupling efficiency\cite{carstensMegawattscaleAveragepowerUltrashort2014}, or change the mode waist at the focal point\cite{luStable500KW2024}.

We present an analytical model for the change in mirror curvature due to mode-induced thermoelastic deformation, and show that this deformation limits the maximum achievable intensity at the focus of the cavity. We then measure the change in mirror radius of curvature as a function of circulating power in a near-concentric \FP{} cavity and demonstrate that it is possible to achieve at least $70\%$ of the intensity limit predicted by our model. The existence of such a limit is relevant to high focal intensity applications of \FP{} cavities, e.g. for use in extreme ultraviolet, x-ray, and gamma light sources\cite{pupezaPassiveOpticalResonators2022,liGeneralizedLongitudinalStrong2023,jacquetFirstProductionXrays2024,deitrickHighbrillianceHighfluxCompact2018,martensDesignOpticalSystem2022}, transmission electron microscopy laser phase plates\cite{axelrodLaserPhasePlate2024}, and optical dipole trapping of molecules\cite{singhDynamicsBuffergasloadedDeep2023}.

\section{Analytical Model} \label{sec:theory}
\subsection{Cavity geometry}
We consider a symmetric \FP{} cavity with identical mirrors of radius of curvature (ROC) $R$, a distance between the mirror surfaces along the optical axis of $L$, and a corresponding cavity stability parameter $g=1-L/R$. The Gaussian mode supported by this cavity has a focal waist ($e^{-2}$ intensity radius) halfway between the mirrors of
\begin{align}
    w_0= \sqrt{ \frac{\lambda L}{2\pi}\sqrt{\frac{1+g}{1-g}} }, \label{eqn:w_0}
\end{align}
while the mode radius on the mirror surfaces is 
\begin{align}
    w_1 &= \sqrt{ \frac{\lambda L}{\pi \sqrt{1-g^2}} }. \label{eqn:w_1}
\end{align}
We assume that $L$ and $\lambda$ are independently tunable parameters. The fundamental Gaussian modes are resonant when 
\begin{align}
    \lambda &= 2 L  \left( n + \frac{\arccos \left( g \right)}{\pi} \right)^{-1},
\end{align}
where $n$ is a non-negative integer.

\subsection{Thermoelastic deformation of mirrors}
The mode heats the surface of the mirrors via absorption, changing their shape. The shape of the mirrors determines the shape of the mode, which determines the distribution of heat absorbed by the mirrors. These effects act together to determine the self-consistent mode shape of the thermoelastically deformed cavity. 

We begin by approximating the change of shape as affecting only the mirror's radius of curvature \cite{winklerHeatingOpticalAbsorption1991,carstensCavityEnhanced196KW2013,carstensMegawattscaleAveragepowerUltrashort2014}, such that 
\begin{equation}
\frac{1}{R} =\frac{1}{R_0}+\frac{1}{R_{\mathrm{th}}}, \label{eqn:1/R}
\end{equation}
where $R_0$ is the radius of curvature of the undeformed mirror and $R_{\rm th}$ accounts for changes due to thermoelastic deformation. If the thermal properties of the mirror substrate are linear, $R_{\mathrm{th}}^{-1}$ must be proportional to the power $Q$ that is absorbed by each mirror surface, and to the thermal distortivity of the mirror substrate 
\be
\delta = \alpha  \frac{1+\nu}{\kappa}, \label{eqn:mirror_distortivity}
\ee
where $\nu$ is Poisson's ratio, $\alpha$ the coefficient of thermal expansion (CTE), and $\kappa$ the thermal conductivity \cite{barberContactMechanics2018}. We assume that the absorbed power $Q= A P$ is proportional to the cavity's circulating power $P$ via an absorption coefficient $A$. Since $w_1$ is the only length scale relevant to the heating of the mirror surface (assuming that the mirror dimensions are much larger than the mode radius on the mirror surface), dimensional analysis then requires that 
\begin{align}
R_{\mathrm{th}}^{-1} &= - M \frac{P}{w_1^2}, \label{eqn:R_th}
\end{align}
where we call the parameter $M = N  A \delta$ the ``mirror distortivity" with $N>0$ a dimensionless constant. The mirror distortivity encapsulates all relevant thermoelastic properties of the mirror in a single parameter.

\subsection{Self-consistent cavity mode} 
We now determine a self-consistent solution by requiring that the modified radius of curvature $R$ supports the mode that heats the mirrors in such a way as to lead to that radius of curvature. 

From equations \eqref{eqn:1/R} and \eqref{eqn:R_th} we write the deformed cavity's stability parameter as a function of the mode's circulating power and radius on the mirror surface:
\begin{align}
    g &= g_0 + M L  \frac{P}{w_1^2},
\end{align}
where $g_0 = 1 - L/R_0$ is the stability parameter of the undeformed cavity. We insert $w_1$ from equation \eqref{eqn:w_1} to obtain the self-consistent solution for the deformed cavity's stability parameter:
\begin{align}
g &= g_0 + \frac{1 - g_0^2}{g_0 \pm \sqrt{1+ \frac{1-g_0^2}{p^2}}}, \label{eqn:g_p}
\end{align}
where $p = \pi \left| M \right| P / \lambda$ is a dimensionless parameterization of the circulating power. Since $w_1$ is real-valued only when $-1 \leq g \leq 1$, all physical solutions of \cref{eqn:g_p} necessarily satisfy this condition.

\begin{figure}[t]
\centering
\includegraphics[width=0.75\textwidth]{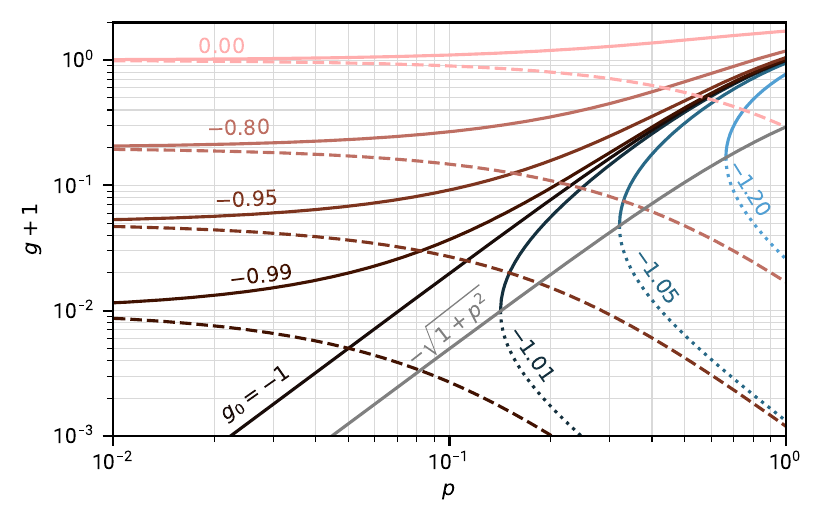}
\caption{\label{fig:g_vs_p} The deformed cavity stability parameter (plus one) $g+1$ (see equation \eqref{eqn:g_p}) as a function of the normalized cavity circulating power $p$, for several different values of $g_0$. Solid lines: $M>0$, ``$+$" branch solutions; dotted lines:  $M>0$, ``$-$" branch solutions; dashed lines: $M<0$, ``$-$" branch solutions. The boundary between the ``$+$" and ``$-$" branch solutions $g_0=-\sqrt{1+p^2}$ is shown in gray.}
\end{figure}

The ``$+$" branch of equation \eqref{eqn:g_p} is valid when $M>0$, $-\sqrt{1+p^2} \leq g_0 \leq 1$, while the ``$-$" branch is valid when $M<0$,  $-1\leq g_0 \leq 1$ or $M>0$,  $-\sqrt{1+p^2} \leq g_0 < -1$. We ignore solutions with $g_0>1$ since they correspond to the unphysical case that $L<0$ under our assumption that $R_0>0$. Equation \eqref{eqn:g_p} is graphed as a function of $p$ for several different values of $g_0$ in \cref{fig:g_vs_p}. Notably, when $M>0$, a solution on each branch exists for $- \sqrt{1+p^2} \leq g_0 < -1$, beyond the normal stability range ($-1\leq g_0 \leq 1$) of the undeformed cavity. In principle, the cavity can be brought to this condition on the ``$+$" branch by introducing circulating power to a geometrically stable cavity with $-1 \leq g_0 \leq 1$ and then increasing the cavity length while maintaining circulating power. However, it is unclear if this can be achieved in practice (section \ref{sec:limitations}). It is also unclear how the ``$-$" branch can be reached since it only connects to the ``$+$" branch when $g_0 = - \sqrt{1+p^2}$ at the edge of the region where valid (real) solutions of equation \eqref{eqn:g_p} exist. For these reasons we mostly restrict our attention to the solutions with $-1 \leq g_0 < 1$.

\subsection{Mode shape and limiting intensity}

Equations \eqref{eqn:w_0} and \eqref{eqn:g_p} yield an expression for the focal waist of the deformed cavity:
\begin{align}\label{eq: deformed w_0}
     w_0 &=   \sqrt{ \frac{\lambda R_0 }{2 \pi}   }  \left(  \left(g_0^2 - 1\right) \frac{ 1 \pm \sqrt{1+\frac{1-g_0^2}{p^2}}  }{  1 \mp \sqrt{1+\frac{1-g_0^2}{p^2}}  } \right)^{1/4}.
\end{align}
Considering the ``+" branch solutions ($M>0$, $-\sqrt{1+p^2} < g_0 \leq 1$), we find that as $p \to \infty$ for any fixed $g_0$, the focal waist approaches 
\begin{align}
    w_{0,\mathrm{min}} &=  \sqrt{ \frac{\lambda R_0 p }{ \pi}   }. \label{eqn:w_0_min}
\end{align}
For the special case of $g_0 = -1$, the focal waist is always given by $w_{0,\mathrm{min}}$.
Interestingly, since this focal waist is proportional to the square root of the circulating power, the corresponding maximum achievable focal intensity is 
\begin{align}
    I_{0,\mathrm{max}} &= \frac{8}{\pi M R_0}, \label{eqn:I_0_max}
\end{align}
which is finite, independent of the circulating power, and only depends on the mirror radius of curvature and mirror distortivity parameter. That is, in the limiting case, an increase in circulating power will increase the focal waist due to thermoelastic deformation of the mirrors, and will not lead to a larger intensity. Note that \cref{eqn:I_0_max} gives the intensity at an antinode of the cavity mode standing wave, which is $8P/\pi w_0^2$ owing to the interference of two counter-propagating beams, each having a power $P$. 

Even if it is possible to achieve the condition $-\sqrt{1+p^2} \leq g_0 < -1$, such limits still exist on the solution boundary $g_0 = -\sqrt{1+p^2}$ for any $p>0$: the minimum achievable focal waist is a factor of $\sqrt{2}$ smaller than that given in equation \eqref{eqn:w_0_min} and therefore the corresponding intensity limit is a factor of $2$ higher than given in equation \eqref{eqn:I_0_max}. No such limits exist if it is possible to operate the cavity on the ``$-$" solution branch in the region $-\sqrt{1+p^2} \leq g_0 < -1$ since those solutions have $g\to -1$ as $p \to \infty$.

If $M<0$ then $g\to -1$ when $p\to \infty$ for any value of $g_0$ for which a solution exists, and so no minimum focal waist or maximum intensity limits exist. Such behavior could be useful in a high-circulating-power near-concentric cavity, since the thermoelastic deformation naturally pushes the cavity towards but not beyond concentricity. $M<0$ may be possible using a mirror substrate material with a negative CTE.

\subsection{Limitations of the model} \label{sec:limitations}
Our model is only valid within the paraxial approximation $w_0 \gg \lambda / \pi$ since equations \eqref{eqn:w_0} and \eqref{eqn:w_1} are only valid in that limit. Using equation \eqref{eqn:w_0}, we find that in terms of the cavity stability parameter the paraxial approximation is satisfied when $g+1 \gg 8 \left(\lambda / \left(\pi L\right)\right)^2$. 

So far, our model does not account for the fact that the thermoelastic curvature change $R_{\rm th}^{-1}$ of the mirror surface will not be uniform over the surface of the mirror but will decrease in magnitude with distance from the center of the cavity mode. This leads to increasingly large aspherical aberration of the cavity mode as the undeformed cavity configuration approaches concentricity ($g_0 \to -1$). While for $g_0<-1$ it is still possible that the center of the deformed cavity mirror is geometrically stable, we hypothesize that this leads to a substantially distorted mode shape and increased diffraction loss from the cavity mode. This may preclude high circulating power and/or small focal waists. Thus, the results derived so far are most accurate when the undistorted cavity is stable,  $-1\leq g_0 \leq 1$. The following section accounts for non-spherical changes to the mirror surface profile to leading order in a perturbative expansion.

\subsection{Analytic expression for the dimensionless constant $N$}
To determine the numerical factor $N$ in the expression for the mirror distortivity parameter $M=N A \delta$ and consider the effect of the aspherical distortion, we now study the mirror surface deformation arising from a Gaussian laser beam. The displacement of the mirror surface in the $z$-direction (defined to be along the optical axis pointing into the mirror) for a point-like source of heat $Q$ applied to the surface is taken from Ref. \cite{barberContactMechanics2018}, equation~(17.20): 
\begin{align}
G\!\left(\vb{r}\right) &= \frac{Q\delta}{2\pi}\ln \left(r\right), \label{Greens}
\end{align}
where $\vb{r} \in \mathbb{R}^2$ is the position on the surface of the mirror, $r = \left|\vb{r}\right|$, and $\delta$ the thermal distortivity as defined in \cref{eqn:mirror_distortivity}. For a Gaussian heat distribution,
\begin{align}
q\!\left(\vb{r}\right) &= \frac{2Q}{\pi w_1^2}e^{-2 r^2/w_1^2},
\end{align}
where $w_1$ is the laser beam radius at $1/e^2$ intensity on the mirror and $Q$ the total dissipated heat, we use \cref{Greens} as a Green's function and integrate to find the axial displacement of the mirror surface
\begin{align}
u\!\left(\vb{r}\right) &= \int q \left( \vb{r}' \right) G\left( \vb{r} - \vb{r}' \right) d^2 \vb{r}' \label{eq: u from greens setup}\\
&= \frac{2Q\delta}{2\pi^2 w_1^2} \int e^{-2r'^2/w_1^2} 
 \ln\left(\sqrt{r^2+r'^2-2rr'\cos\theta}\right)r' dr' d\theta' \nonumber \label{distortion} \\
&= \frac{Q\delta}{\pi^2}\left[\frac \pi 2\ln\left(\frac{r}{w_1}\right)-\frac \pi 4 \Ei \left(-2\frac{r^2}{w_1^2}\right)\right] \\
&= \frac{Q\delta}{\pi^2}\left(-\frac \pi 4 [\Gamma +\ln(2) ]+\frac \pi 2 \frac{r^2}{w_1^2}+\cdots \right), \label{deform1}
\end{align}
where $\Gamma$ is the Euler-Mascheroni constant, and $\Ei$ the exponential-integral function. The corresponding temperature distribution is
\begin{align}
T\!\left(\vb{r}\right) &= \frac{Q}{\sqrt{2\pi} \kappa w_1} \exp\!\left(-\frac{r^2}{w_1^2}\right) \mathcal{I}_0\!\left(\frac{r^2}{w_1^2}\right), \label{eq:temperature}
\end{align}
where $\mathcal{I}_0$ is the modified Bessel function of the first kind of zeroth order.

\begin{figure}[t]
\centering
\includegraphics[width=0.75\textwidth]{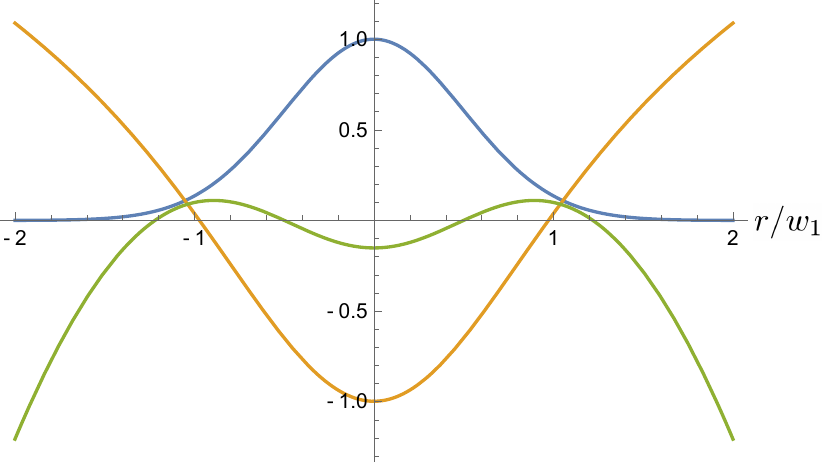}
\caption{\label{Deformation} Mirror deformation and beam intensity. Blue: Gaussian beam intensity (arbitrary units), Orange: Deformation of the mirror surface, \cref{distortion}. Green: Aspherical part of the deformation, \cref{deform2}, assuming $\beta = 1/2$.}
\end{figure}

\Cref{Deformation} shows the mirror surface figure distortion calculated in equation \eqref{deform1}. It is parabolic near the center of the mirror, which amounts to a change in the radius of curvature of the mirror in the paraxial approximation. Farther away from the center, however, the distortion deviates from spherical. Our strategy to treat the effects of this deformation is as follows: We rewrite the deformation $u\!\left(\vb{r}\right)$ as the sum of a spherical part $s\!\left( \vb{r}\right)$ and a residual aspherical part $a\!\left(\vb{r}\right)$. 
\begin{align}
u\!\left(\vb{r}\right) &= s\!\left(\vb{r}\right) + a\!\left(\vb{r}\right) \\
s\!\left(\vb{r}\right) &= \frac{Q\delta}{\pi^2}\left(\gamma + \beta \frac{\pi}{2}  \frac{r^2}{w_1^2}\right) \label{deform2},
\end{align}
The spherical part is the sum of a constant term $Q\delta\gamma /\pi^2$ (which we absorb into our definition of the cavity length $L$) and a quadratic term $Q\delta \beta r^2/\left(2\pi w_1^2 \right)$. This decomposition is not an approximation, and the two free dimensionless parameters $\gamma$ and $\beta$ are, in principle, arbitrary. However, we show in Appendix \ref{Perturbation} that $\beta =1/2$ and $\gamma = -\pi (1+2\Gamma)/8 \approx -0.846$ minimize the magnitude of aspherical perturbation terms. Specifically, the perturbation proportional to a Laguerre-Gaussian mode with radial index $\rho=1$ vanishes with this choice, leaving only the small contributions of $\rho>1$ modes. Of course, the actual shape of the cavity mode is independent of our choice, but we are free to express it in a set of Laguerre-Gaussian modes that contains the fewest non-negligible terms.

The spherical part amounts to a change in the radius of curvature given by
\be
R_{\rm th}^{-1} = - \frac{\beta Q\delta}{\pi w_1^2}.
\ee
Comparing to \cref{eqn:R_th} and noting that $Q = A P$, we find the coefficient
\begin{align}\label{eq: N analytic}
N = \frac{\beta}{\pi} = \frac{1}{2\pi}.
\end{align}
As a point of comparison, we are also able to non-perturbatively simulate the aberrated optical cavity eigenmodes resulting from the full displacement profile $u\!\left(\vb{r} \right)$ using the method of~\cite{benedikterTransversemodeCouplingDiffraction2015}. The aberrated eigenmode leads to a non-Gaussian heat distribution, which modifies $u\!\left(\vb{r} \right)$ through \cref{eq: u from greens setup}, and feeds back to the aberrated mode profile. Using an iterative numerical procedure, the self-consistent cavity eigenmode and deformed mirror profile can be determined. The numerical results agree with \cref{eq: deformed w_0}, with a coefficient $N = (1\pm0.1)/2\pi$, consistent with \cref{eq: N analytic}. Thus, with our choice of $\beta$ and $\gamma$ to define the spherical part of the mirror surface deformation, almost all of the effects of the deformation are well explained by a change to the mirror radius of curvature.

The remaining aspherical part of the deformation is given by
\begin{align}
a\!\left(\vb{r}\right)
= \frac{Q\delta}{\pi^2}\left[\frac \pi 2\ln\left(\frac{r}{w_1}\right)-\frac \pi 4 \Ei \left(-2\frac{r^2}{w_1^2}\right)-\gamma - \frac{\pi \beta r^2}{2 w_1^2}\right] . \label{aspher}
\end{align}
The effect of this aspherical part is explored in \cref{Perturbation}. At low powers, we find the aspherical corrections to the intensity to scale with $p$, so the spherical part explains the vast majority of the deformation. When $p\gg 1$, the aspherical corrections produce a $\approx5\%$ adjustment to the intensity limit \cref{eqn:I_0_max},
\begin{align}
I_{0, \mathrm{max}}' =  \left[\frac 32 +{\rm Li}_2\left(-\frac 12\right )\right] I_{0, \mathrm{max}} \approx 1680 \, \frac{\rm GW}{{\rm cm}^2}\left(\frac{R_0}{\rm cm}\,\frac{A}{{10^{-6}}}\,\frac{\delta}{100\,{\rm nm/W}}\right)^{-1} \label{eqn:I_0_max_aspherical},
\end{align}
which does not account for the (very small) change to the aspherical deformation due to non-Gaussian corrections to the aberrated mode profile.

\section{Experimental Results}
We have measured the mirror distortivity in two near-concentric \FP{} cavities which use Corning ULE~7972 substrate mirrors with a nominal ROC of $R_0 = \qty{10}{\milli\meter}$. The mirrors have ion-beam-sputtered dielectric coatings with a nominal transmission of $\qty{80e-6}{}$ at the laser wavelength of $\lambda \approx \qty{1064}{\nano\meter}$. The design of these cavities is described in detail in \cite{axelrodLaserPhasePlate2024}. The cavities have an identical geometry and coating design, but differ in the mirror absorption $A$. Both cavities are operated in the near-concentric regime with $g_0 \approx -0.9996$.

We measure the change in mirror ROC by measuring the cavity's transverse mode spectrum and laser wavelength. The frequencies of the (first-order) transverse modes are translated to the corresponding cavity stability parameter (using equation~(2.37) of \cite{axelrodLaserPhasePlate2024}). Changes in the cavity length during the experiment due to thermal expansion of the cavity mount when the circulating power is changed are inferred from the measured changes in the laser wavelength. An initial cavity length of $\qty{20}{\milli\meter}$ is assumed since the results of the measurement only depend on changes in the cavity length around this nominal value. The mirror radius of curvature can then be calculated from the cavity stability factor and cavity length measurement (see equation~(2.21) of \cite{axelrodLaserPhasePlate2024}). This procedure is applied to both the $\left(1,0\right)$ and $\left(0,1\right)$ TEM cavity modes, and the resulting radii of curvature averaged (geometric mean).

This data is then fit with the model described by equation \eqref{eqn:g_p}, where the laser wavelength $\lambda$, cavity length $L$, and circulating power $P$ are measured for each data point, and the zero-circulating power mirror radius of curvature $R_0$ (equivalent to the zero-circulating power stability parameter $g_0$) and the mirror distortivity parameter $M$ are used as fit parameters.

\subsection{High-absorption cavity}
\begin{figure}[t]
	\centering
    \includegraphics[width=2.5in]{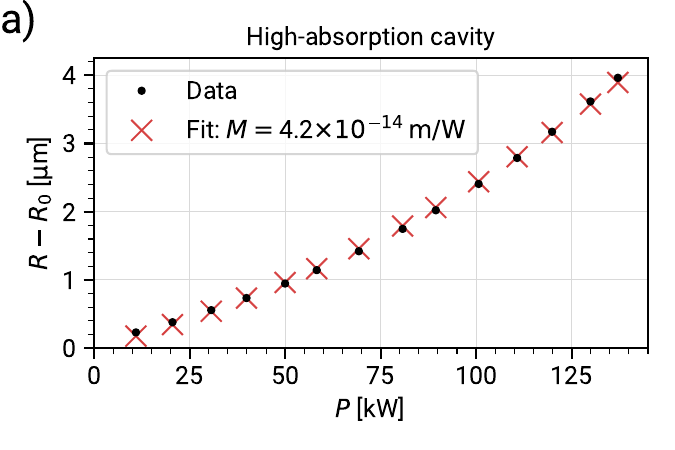}
    \hfill
    \includegraphics[width=2.5in]{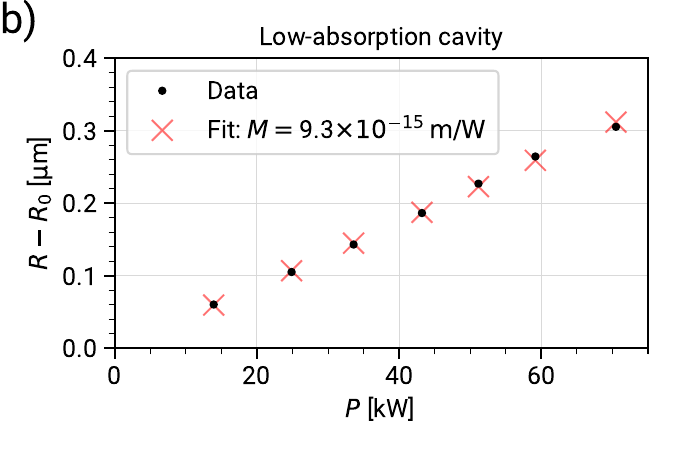}
	\caption{Change in the radius of curvature of the mirrors $R-R_0$ as a function of circulating power $P$ in the a) high-absorption and b) low-absorption cavities. Black dots show the measured data, red X's show the best fit to the model described by \cref{eqn:g_p}.
    }
	\label{fig:mirror_ROC_change}
\end{figure}
While both cavities have mirror coatings that are expected to have absorption coefficients around $A=\qty{1e-6}{}$, one of them suffered slight contamination during assembly and has a higher absorption than usual. The results of the measurement for this cavity are shown in \cref{fig:mirror_ROC_change}a. The fit shows reasonable agreement with the data, though there is a small systematic error in the residuals which causes the fit to underestimate the radius of curvature change at low and high circulating powers. We have not yet determined the source of this error. One possibility is that it is due to the temperature dependence of the ULE~7972 mirror substrates' CTE \cite{ULECorningCode2016}.
It could also be from the model approximating the change in mirror surface due to thermoelastic deformation as being simply a change in radius of curvature.

\begin{figure}[t]
	\centering
	\includegraphics[width=3.5in]{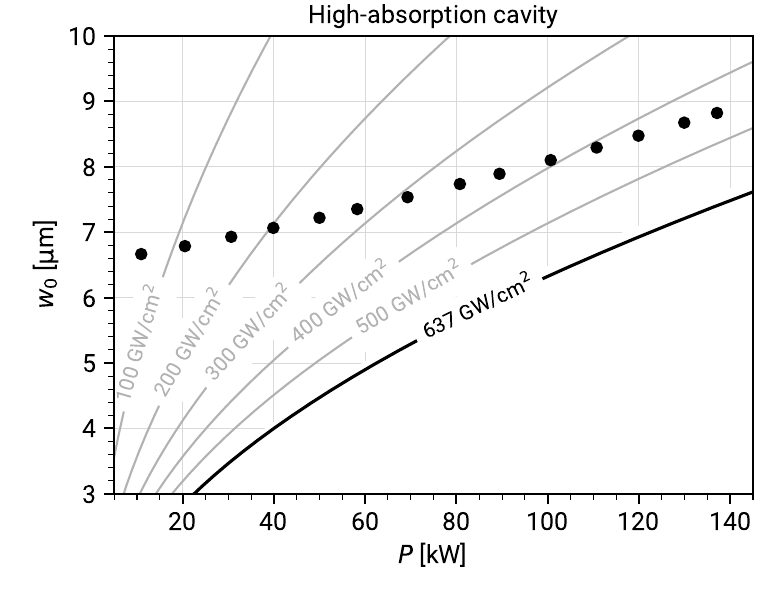}
	\caption{Measurements of the mode focal waist as a function of circulating power (black dots) using the high-absorption cavity. The corresponding standing-wave focal intensities are shown with gray lines. The predicted maximum focal intensity is shown with the black line.
    }
	\label{fig:intensity_limit_saturation_gen_3_2}
\end{figure}

As expected, the data shows an increase in mirror radius of curvature with increasing circulating power. The nonlinearity of the increase is due to the lower numerical aperture mode (at higher radius of curvature) causing yet more thermal deformation of the mirror substrate. The fit model has a mirror distortivity parameter of $M = \qty{4.2e-14}{\meter/\watt}$. According to \cref{eqn:I_0_max_aspherical}, this should restrict the cavity focal intensity to a maximum of $I_{0,\mathrm{max}}' = \qty{637}{\giga\watt/\centi\meter^{2}}$. \Cref{fig:intensity_limit_saturation_gen_3_2} shows the cavity mode focal waist $w_0$ as a function of circulating power from the data used in \cref{fig:mirror_ROC_change}a.
This illustrates how the cavity achieved a focal intensity of $\qty{450}{\giga\watt/\centi\meter^2}$, which is $70\%$ of the predicted limit. Note that the maximum value of $p$ achieved in this experiment was only $0.017$. 

Assuming the thermal distortivity to be the temperature-averaged (\qty{20}{\celsius}--\qty{150}{\celsius}) nominal value $\delta = \qty{45}{\nano\meter/\watt}$ of ULE~7972 \cite{ULECorningCode2016} gives an absorption coefficient of $A=5.8 \times 10^{-6}$, which implies an implausibly high peak temperature on the mirror surfaces ($\sim\qty{900}{K}$, see \cref{eq:temperature}). It is plausible that the effective thermal distortivity may be higher than the nominal substrate value because of contributions from the coating, both directly from its higher CTE and indirectly through thermally-induced stress because of the CTE mismatch. We have not investigated this in detail, but note that the data are seemingly well-described by a single effective thermal distortivity parameter. This distortivity can be determined from the data if combined with an independent, ideally in situ measurement of the absorption coefficient.

\subsection{Low-absorption cavity}
The results of the measurement for the low-absorption cavity are shown in \cref{fig:mirror_ROC_change}b. The fitted mirror distortivity of $M = \qty{9.3e-15}{\meter/\watt}$, 4.5 times lower than for the high-absorption cavity, corresponds to a predicted focal intensity limit of $\qty{2.9}{\tera\watt/\centi\meter^{2}}$.

Combining this value for the mirror distortivity with the nominal thermal distortivity $\delta = \qty{45}{\nano\meter/\watt}$ of ULE~7972 implies an absorption coefficient of $A= \qty{1.3e-6}{}$.
This is several times larger than the absorption of $\qty{0.4e-6}{}$ measured by photothermal common-path interferometry~\cite{alexandrovskiPhotothermalCommonpathInterferometry2009} on a witness sample coated in the same coating run as the mirrors in this cavity. If we instead assume that the witness sample measured absorption accurately represents the absorption of the cavity mirrors, the effective thermal distortivity is $\delta = \qty{146}{\nano\meter/\watt}$, which suggests a coating contribution to the distortivity on the order of $\qty{100}{\nano\meter/\watt}$.

\section{Discussion}
Our analytical model predicts that thermoelastic deformation of the mirrors in a \FP{} cavity sets an upper limit on the focal intensity. Our experimental results show that it is possible to achieve a substantial fraction ($70\%$) of this intensity limit, and that the predicted limit can be as high as $\qty{2.9}{\tera\watt/\centi\meter^{2}}$ using low-absorption ion-beam sputtered dielectric mirrors with Corning ULE~7972 substrates.

It should be possible to further reduce the effects of thermoelastic deformation and increase the predicted intensity limit by using mirrors with a smaller radius of curvature $R_0$, smaller absorption $A$, and/or smaller (positive) thermal distortivity $\delta$. It is possible to manufacture cavity mirrors capable of supporting high circulating powers (few/small enough surface defects to avoid laser-induced damage, and low enough surface roughness to support a high cavity finesse) with a ROC of $R_0 = \qty{5}{\milli\meter}$ or perhaps somewhat lower using conventional polishing methods. Smaller ROCs on the order of $\qty{100}{\micro\meter}$ can be achieved in fiber tip cavities\cite{pfeiferAchievementsPerspectivesOptical2022} or micro-fabricated mirrors\cite{jinMicrofabricatedMirrorsFinesse2022a}. However, a smaller ROC also leads to a higher intensity on the mirror surfaces (for the same focal intensity), which may cause laser-induced damage of the mirrors and limit the focal intensity to well below that predicted by equation \eqref{eqn:I_0_max_aspherical}.

Our model indicates that mirrors with a negative CTE such that $M<0$ may be the most promising approach to avoid a thermoelastic deformation focal intensity limit and simultaneously achieve a small focal waist and high circulating power in a \FP{} cavity. Examples of conventional optical substrate materials with negative CTE include Corning ULE~7972 at a temperature below $\qty{295}{\kelvin}$, fused silica at a temperature below $\qty{170}{\kelvin}$, and silicon at a temperature between $\qty{20}{\kelvin}$ and $\qty{120}{\kelvin}$ \cite{axelrodLaserPhasePlate2024}.
Alternatively, substrates with small average CTEs over a wide temperature range (such as Schott Zerodur Expansion Class 0 Special) may be used \cite{maisenbacherUltrahighContinuouswaveIntensities2026}. However, we note that coating contributions to the effective thermal distortivity must be taken into account, as they may dominate over the substrate's contribution. In this case, it should still be possible to reduce the magnitude of the effective thermal distortivity by using a substrate material with higher thermal conductivity (e.g., diamond or cryogenic sapphire), thereby reducing the temperature change and concomitant thermoelastic deformation in the coating \cite{axelrodLaserPhasePlate2024}.

Cavities with more than two mirrors can be engineered such that thermoelastic deformation of the mirrors with $M>0$ increases the mode waist at the mirror surfaces, thereby avoiding the type of focal intensity limit predicted by our model (similar to the case of a \FP{} cavity with $M<0$) \cite{carstensLargemodeEnhancementCavities2013}. However, using more mirrors decreases the maximum possible finesse of the cavity and increases its size and optomechanical complexity.

\begin{backmatter}
\bmsection{Funding}
Brown Science Foundation through the Brown Investigator Award (award no. 1167); Chan Zuckerberg Initiative (award no. 2021-234606 and 2025-367757); Gordon and Betty Moore Foundation (grant no. 9366); Heising-Simons Foundation (grant no. 2023-4467); National Institutes of Health grant R01GM126011; U.S. Department of Energy, Office of Science, National Quantum Information Science Research Centers, Quantum Systems Accelerator (award no. DESCL0000121); Lawrence Berkeley National Laboratory Directed Research and Development Program; Cooperative Research and Development Agreement award AWD00004352.

\bmsection{Acknowledgments}
The authors thank Osip Schwartz for helpful discussions on the properties of the self-consistent cavity mode solutions, and Carter Turnbaugh for assisting in preliminary data collection. LM acknowledges a Feodor Lynen Fellowship from the Alexander von Humboldt Foundation. AS acknowledges support from the UC Berkeley Physics Department’s Graduate Student Support Fund. PNP acknowledges National Institutes of Health fellowship F32GM149186.

\bmsection{Disclosures}
\noindent (P) JJA, HM are inventors on US patents US10395888B2, US11990313B2.

\bmsection{Data availability} Data underlying the results presented in this paper are not publicly available at this time but may be obtained from the authors upon reasonable request.

\end{backmatter}

%%%%%%%%%% If using BibTeX:
\bibliography{Thermoelastic_deformation_paper} % Export this BibTeX file from Zotero

\begin{appendix}
\section{Cavity mode with aspherically deformed mirrors}\label{Perturbation}

We now calculate the distortion of the cavity mode that arises from the aspherical part of the mirror distortion, using the (Gaussian) self-consistent mode as a starting point. Of course, one could go further and consider a correction to the distortion based on the corrected, non-Gaussian mode shape. We will, however, find that the corrections calculated in this section are already small, so that corrections of even higher order will be negligible. 

We start by expanding the electric field of the cavity mode $E\left(r,\phi,z\right)$ in Laguerre-Gaussian modes LG$_{\Lambda \rho}$, where $\Lambda=0$ because of the cylindrical symmetry. We express these modes using a $z-$coordinate which is zero at the center of the cavity and $z= \pm L/2 \equiv \pm \ell$ at the mirrors. Thus, 
\begin{align} \label{cylsymm}
E\left(r,\phi,z\right)  = & \sqrt{\frac 2\pi} \frac{w_0}{w(z)}\exp\left(-\frac{r^2}{w(z)^2}\right) \exp\left(-ikz -ik\frac{r^2}{2R(z)} +i\psi(z)\right) \nonumber \\ & \cdot \left[1+\sum_\rho \alpha_\rho L_\rho\left(\frac{2r^2}{w^2(z)}\right) \exp\left( i 2 \rho \psi(z)\right) \right].
\end{align}
where $k = 2 \pi /\lambda$, $R(z)$ is the wavefront curvature and $\psi(z)$ the Gouy phase. The $\alpha_\rho$ are the expansion coefficients that we wish to determine. Note that the symmetry of the cavity demands $\alpha_\rho \in \mathbb{R}$. If the $|\alpha_\rho|\ll 1$, the final bracket approximates an exponential function:
\begin{align}
E \left(r,\phi,z\right) = & \sqrt{\frac 2\pi} \frac{w_0}{w(z)}\exp\left(-\frac{r^2}{w(z)^2}\right) \exp\left(-ikz -ik\frac{r^2}{2R(z)} +i\psi(z)\right) \nonumber \\ & \cdot  \exp\left[\sum_\rho \alpha_\rho L_\rho\left(\frac{2r^2}{w^2(z)}\right) \exp\left( i 2 \rho \psi(z)\right) \right].
\end{align}
This mode must match the boundary condition that there be no parallel electric field on the mirror surface. This can be achieved by setting the imaginary part of the combined exponents to zero. The first exponent, $-ikz -ikr^2/2R(z)+i\psi(z)$, vanishes at the surface of the mirror before it is affected by the aspherical part of the distortion. Thus, the second exponent must equal the aspherical part of the distortion $a\left(\vb{r}\right)$,
\begin{align}
a\left(\vb{r}\right) = \frac 1k \sum_{\rho \geq 1}  L_\rho\left(\frac{2r^2}{w^2(z)}\right) \Imr\left( \alpha_\rho e^{i 2 \rho \psi(z)}\right). 
\end{align}
To solve for the $\alpha_\rho$, we multiply both sides with a Gaussian function times a Laguerre polynomial, and integrate over $r$, using the orthogonality of the Laguerre polynomials:
\begin{align}
&\int_0^\infty a\left(\vb{r}\right) L_q\left(\frac{2r^2}{w^2}\right) e^{-2r^2/w^2(z)} r dr \nonumber \\
&\quad =\frac 1k \int_0^\infty \sum_{\rho \geq 1}  \Imr\left( \alpha_\rho e^{i 2 \rho \psi(z)}\right) L_\rho \left(\frac{2r^2}{w^2}\right) L_q\left(\frac{2r^2}{w^2}\right) e^{-2r^2/w^2} r dr \nonumber \\
\Rightarrow \quad & \frac{4k}{w^2} \int_0^\infty a\left(\vb{r}\right)  L_q\left(\frac{2r^2}{w^2}\right) e^{-2r^2/w^2} r dr  =   \Imr \left( \alpha_q e^{i 2 q \psi(z)}\right).
\end{align}
We can now insert $a\left(\vb{r}\right)$ from equation \eqref{aspher} and substitute $r' = r/w$
\be\label{alphaeq}
4k\frac{Q\delta}{\pi^2}[C_q + D_q]  = \Imr\left( \alpha_q e^{i 2 q \psi(z)}\right) 
\ee
where
\bea
C_q&=&\int_0^\infty \left[\frac \pi 2\ln\left(r'\right)-\frac \pi 4 {\rm Ei}\left(-2 r'^2\right)\right] e^{-2r'^2} L_q\left(2 r'^2\right) r'  dr' \nonumber \\
D_q&=& \int_0^\infty \left[-\gamma - \frac\pi 2 \beta r'^2 \right] e^{-2 r'^2} L_q\left(2 r'^2\right) r'  dr'.
\eea
We find
\be
C_0 = -\frac{\pi\Gamma }{16}, \quad C_{q>0} = -\frac{\pi}{32}\frac{1}{2^{q-1}q}, \quad D_0=\frac{ -\pi\beta-4\gamma}{16}, \quad D_1 = \frac{\pi \beta}{16}, \quad D_{q>1}=0.
\ee
Thus, if we choose $\beta=1/2$ and $\gamma=-\pi(1 + 2\Gamma)/8 $, then the (usually dominant) lowest-order Laguerre-Gaussian corrections $C_0+D_0 = C_1+D_1=0$. Thus, we need to consider only the (small) terms with $q= 2,3, \ldots $.  

\subsection{Some properties of the distorted cavity mode}
The power in each mode is proportional to the integral of the intensity over the beam area and is easiest evaluated at $z=0$:
\be
P_q \propto \left|\alpha_q\right|^2\frac 2\pi \int_0^\infty e^{-2r^2/w_0^2}2 \pi rdr =  \left|\alpha_q\right|^2 w_0^2.
\ee
When the mirror is distorted, power is transferred from the fundamental mode to higher order modes. The fraction of the transferred power is 
\be
\frac{P_{q>0}}{P_0}  =\sum_{q \geq 2} \left|\alpha_q\right|^2,
\ee
The intensity $I$ of the distorted beam at the origin $z=0, r=0$ relative to the intensity $I_0$ of the undistorted beam is given by 
\be
\frac{I}{I_0} = \left|1+\sum_{q\geq 2} \alpha_q \right|^2
\ee

\subsection{Near-concentric cavity}\label{nearconcentric}
Our beam waist is located at $z=0$ while the mirrors are at $z=\pm L/2 = \pm \ell$. The Gouy phase is 
\begin{align}
    \psi=\psi(\ell) = \arctan\left(\frac{\ell}{z_R}\right)=\frac \pi 2-\arctan\left(\frac{z_R}{\ell}\right),
\end{align} 
where $z_R=\pi w_0^2/\lambda$ is the Rayleigh range. In a near-concentric cavity, $\ell \gg z_R$ and the Gouy phase is $\psi \approx \pi/2-z_R/\ell$.  Within the same approximation, $\sin(2q\psi) \approx \sin(q \pi -2q z_R/\ell) = -(-1)^q \sin(2qz_R/\ell)$, so \cref{alphaeq} yields
\be
\alpha_q \approx 
 \frac{kQ\ell\delta}{8\pi z_R}\frac{(-1)^q}{2^q q^2} = B\frac{(-1)^q}{2^q q^2} 
\ee
where 
\begin{align}
    B=\frac{k Q \ell \delta}{8\pi z_R} =\frac{Q \ell \delta}{4\pi w_0^2} \approx \frac{Q \delta }{4 \lambda }\sqrt{-\frac{g_0}{1+g_0+p^2}}
\end{align}
where we used the expression for the waist $w_0^2=(\ell \lambda/\pi) \sqrt{-(1+g_0+p^2)/g_0}$ of the self consistent mode for $g\approx -1$. In the low-power limit $p\ll \sqrt{1+g_0}$, the leading order contribution is $B = Q\delta\sqrt{-g_0/(1+g_0)}/(4\lambda)$, which scales linearly with $p$, since $Q\propto p$. On the other hand, when $p$ is large, we find (assuming $\delta>0$)
\begin{align}
    B\rightarrow \frac{Q \delta }{4 \lambda p } = \frac 12. 
\end{align}
For the intensity at the origin, we obtain 
\begin{align}
\frac{I}{I_0} = \left|1+B\left(\sum_{q=1}^\infty \frac{(-1)^q}{2^qq^2} +\frac12\right)\right|^2 = 1+2B\left[{\rm Li}_2\left(-\frac 12\right)+\frac 12\right] + \ldots \rightarrow 1.051
\end{align}
where the ellipsis indicates terms of order $B^2$. Since $B\propto p$ for low $p$, the corrections to the intensity due to the aspherical mirror distortion are only important when $p\gg \sqrt{1+g_0}$, where they lead to a slight increase of the intensity compared to the case of only spherically-distorted mirrors.

For $p\gg \sqrt{1+g_0}$, the total power transferred into a higher-order Gaussian mode is 
\be\label{phigher}
\frac{P_{q>0}}{P_0}  = \frac 14 \left(\sum_{q=1}^\infty \frac{(1/4)^q}{q^4}-\frac{1}{4}\right) = \frac 14 \left[{\rm Li}_4\left (\frac 14 \right)-\frac{1}{4}\right] \approx 0.00103,
\ee
where Li$_s(z)=\sum_{j=1}^\infty z^j/j^s$ is the polylogarithm function.

\end{appendix}

\end{document}